\documentclass[aps,twocolumn,superscriptaddress]{revtex4}
\usepackage{amsmath}
\usepackage{amsfonts}
\usepackage{mathrsfs}
\usepackage{graphicx}
\usepackage{times}
\usepackage{color}
\UseRawInputEncoding

\def\be{\begin{equation}}
\def\ee{\end{equation}}
\def\bea{\begin{eqnarray}}
\def\eea{\end{eqnarray}}

\begin{document}
\title{Topological flat band with higher winding number in a superradiance lattice}

\author{Shuai Li}
\thanks{These authors contribute equally to this work.}
\affiliation{Ministry of Education Key Laboratory for Nonequilibrium Synthesis and Modulation of Condensed Matter,Shaanxi Province Key Laboratory of Quantum Information and Quantum Optoelectronic Devices, School of Physics, Xi'an Jiaotong University, Xi'an 710049, China}

\author{Rui Tian}
\thanks{These authors contribute equally to this work.}
\affiliation{Ministry of Education Key Laboratory for Nonequilibrium Synthesis and Modulation of Condensed Matter,Shaanxi Province Key Laboratory of Quantum Information and Quantum Optoelectronic Devices, School of Physics, Xi'an Jiaotong University, Xi'an 710049, China}

\author{Huan Wang}
\affiliation{Ministry of Education Key Laboratory for Nonequilibrium Synthesis and Modulation of Condensed Matter,Shaanxi Province Key Laboratory of Quantum Information and Quantum Optoelectronic Devices, School of Physics, Xi'an Jiaotong University, Xi'an 710049, China}

\author{Min Liu}
\affiliation{Ministry of Education Key Laboratory for Nonequilibrium Synthesis and Modulation of Condensed Matter,Shaanxi Province Key Laboratory of Quantum Information and Quantum Optoelectronic Devices, School of Physics, Xi'an Jiaotong University, Xi'an 710049, China}

\author{Liangchao Chen}
\affiliation{State Key Laboratory of Quantum Optics and Quantum Optics Devices,
Institute of Opto-electronics, Shanxi University, Taiyuan, Shanxi 030006, People¡¯s Republic of China}

\author{Bo Liu}
\email{liubophy@gmail.com}
\affiliation{Ministry of Education Key Laboratory for Nonequilibrium Synthesis and Modulation of Condensed Matter,Shaanxi Province Key Laboratory of Quantum Information and Quantum Optoelectronic Devices, School of Physics, Xi'an Jiaotong University, Xi'an 710049, China}
	
\begin{abstract}
A five-level M-type scheme in atomic ensembles is proposed to generate a one-dimensional bipartite superradiance
lattice in momentum space. By taking advantage of this tunable atomic system, we show that various types of Su-Schrieffer-Heeger (SSH) model, including the standard SSH and extended SSH model, can be realized.
Interestingly, it is shown that through changing the Rabi frequencies and detunings in our proposed scheme,
there is a topological phase transition from topological trivial regime with winding number being 0 to topological non-trivial regime with winding number being 2. Furthermore, a robust flat band with higher winding number (being 2) can be achieved in the above topological non-trivial regime, where the superradiance spectra can be utilized as a tool for experimental detection. Our proposal would provide a promising approach to explore new physics, such as fractional topological phases, in the flat bands with higher topological number.
\end{abstract}
	
\maketitle

\section{Introduction}
There has been a surge of interest in flat band physics \cite{leykam2018artificial}, where one or more dispersionless bands exist throughout the Brillouin zone. Many theoretical proposals for searching flat band systems have been made \cite{sutherland1986localization,lieb1989two,mielke1991ferromagnetic,mielke1991ferromagnetism,tasaki1992ferromagnetism,mielke1999ferromagnetism,vidal2000interaction,tasaki2008hubbard,springer2020topological} and its captivation has become exceptionally pronounced following the experimental realization in twisted bilayer graphene \cite{cao2018unconventional,cao2018correlated,yankowitz2019tuning,park2021tunable}. Due to the macroscopic level degeneracy in flat bands, lots of interesting physical phenomena, such as ferromagnetism \cite{lieb1989two,mielke1999ferromagnetism,tasaki2008hubbard,costa2016ferromagnetism,derzhko2010low},
Wigner crystals \cite{wu2007flat} and superconductivity \cite{julku2016geometric,lothman2017universal}, can be induced. In particular, isolated flat bands with non-trivial topological properties have also attracted much attention, since fractional topological phases, such as fractional quantum Hall and fractional Chern insulator states \cite{tang2011high,sun2011nearly,neupert2011fractional,sheng2011fractional,wang2011nearly,xiao2011interface,wang2011fractional,zhao2012quantum,jaworowski2015fractional}, can be simulated without Landau levels. More interestingly, the flat bands with higher topological number (e.g., higher Chern number) can host qualitatively new phases of matter with no analogue in the flat band being similar to the continuum Landau level \cite{trescher2012flat,yang2012topological,liu2012fractional}. New types of intriguing fractional Chern insulator states, for fermions at $\nu=1/2N+1$ and for bosons at $\nu =1/N+1$, are unveiled \cite{liu2012fractional}. Distinguished from the cases in 2D, 1D topological nontrivial flat bands can unusually lead to new physical phenomena, for instance, a charge density wave with a nontrivial Berry phase \cite{budich2013fractional,guo2012fractional}, which is not a 1D analog of the 2D fractional quantum Hall state. However, most previous 1D studies have focused on the flat bands with a unit winding number. To explore the new physics associated with higher winding number flat bands remains unclear and stands as an obstacle to explore.

Here we report the discovery of a new mechanism to achieve the topological flat band with higher winding number in a superradiance lattice. We shall introduce this with a specific model of ultracold atoms, to be illustrated below. The key idea here is to design the non-trivial long-range hopping of atoms in momentum space through our proposed five-level M-type scheme.
Surprisingly, it is shown that through tuning the intensities of the coupling fields, the tunneling between atoms in momentum space is highly tunable and the flat band with higher winding number can be achieved. This idea is motivated by the recent experimental progresses in developing the momentum space lattice composed by the timed Dicke states, i.e., superradiance lattices \cite{wang2015superradiance,chen2018experimental,wang2020synthesized,mi2021time,cai2019experimental,he2021flat,mao2022measuring}, which are the collective atomic excitations with phase correlations. Such phase correlations can be recognized as the momenta of the collective excitations. When they satisfy the phase-matching condition, there are directional superradiant light emissions, which can be utilized as one of the remarkable advantages to explore interesting physics in superradiance lattices, such as chiral current \cite{wang2020synthesized,cai2019experimental}, flat band localization \cite{he2021flat}, and floquet physics \cite{xu2022floquet}. As we shall show with the model below, our proposed five-level M-type scheme can lead to the flat band with higher winding number.

\section{Effective model}
\begin{figure}[b]
	\centering
	\includegraphics[width=0.48\textwidth]{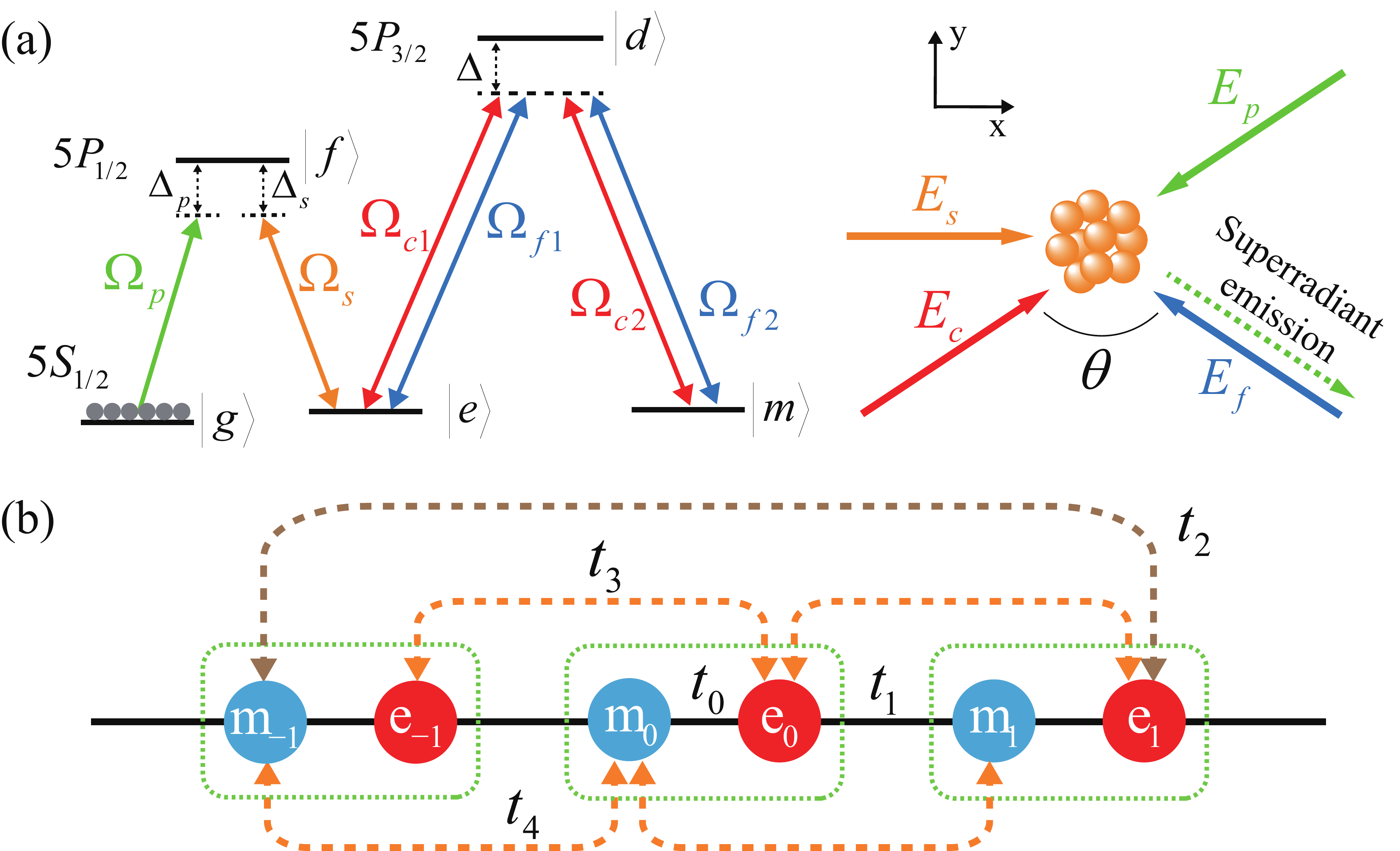}
	\caption{
		(a) The schematic plot of our proposed five-level M-type scheme. The coupling fields $\mathbf{E_c}$ and $\mathbf{E_f}$ drive the transition between $|e\rangle$ and $|m\rangle$. The direction of the $x$-axis is determined by $\mathbf{k_c-k_f}$. The superradiant beam is marked by the dashed line. (b) The 1D bipartite superradiance lattice in momentum space. $t_0$ and $t_1$ stands for the hopping between nearest-neighbor sites within and inbetween the unit cells. $t_2$, $t_3$ and $t_4$ describe the distinct long-range hopping amplitudes, respectively. When varying the Rabi frequencies and detunings in our proposed M-type scheme, all the hopping amplitudes can be changed. Various types of Su-Schrieffer-Heeger (SSH) model, including the standard SSH and extended SSH model, can thus be achieved.}
\label{fig setup}
\end{figure}

Let us take $^{87}Rb$ atomic system as an example to show our proposed five-level M-type scheme, which is schematically presented in Fig.~\ref{fig setup}.
There are two excited states $|f\rangle=|F¡ä = 1,m_F = 0\rangle$, $|d\rangle= |F¡ä¡ä = 1,m_F = 1\rangle$,  which can be selected from $5P_{1/2}$ state and $5P_{3/2}$ state, respectively. And $|g\rangle$, $|e\rangle$, $|m\rangle$ can be chosen from $5S_{1/2}$ state, such as $|F = 1,m_F = 0\rangle$, $|F = 2,m_F = 0\rangle$ and $| F=2,m_F = 2\rangle$.
The probe field $\mathbf{E_p}$ and signal field $\mathbf{E_s}$ couple the states $|g\rangle$ and $|f\rangle$, $|e\rangle$ and $|f\rangle$, respectively. $\Delta_p$ and $\Delta_s$ denote the corresponding frequency detunings. There are two far off-resonant coupling fields $\mathbf{E_c}$ and $\mathbf{E_f}$, where $\mathbf{E_c}$ spontaneously drives both transitions between $|e\rangle$ and $|d\rangle$, $|m\rangle$ and $|d\rangle$, since here $|e\rangle$ and $|m\rangle$ are assumed to be degenerated and the corresponding Rabi frequencies are labeled by $\Omega_{c_1}$ and $\Omega_{c_2}$, respectively. $\mathbf{E_f}$ also spontaneously drives both transitions between $|e\rangle$ and $|d\rangle$, $|m\rangle$ and $|d\rangle$, where the corresponding Rabi frequencies are defined as $\Omega_{f_1}$ and $\Omega_{f_2}$, respectively. $\Delta$ describes the frequency detuning of the above transitions.
$\mathbf{k_{p(s,c,f)}}$ are the wave vectors of the corresponding fields, respectively.
Since $\Delta$ is much larger than other detunings and Rabi frequencies, we can adiabatically eliminate the state $|d\rangle$ and obtain the effective Hamiltonian in the interaction picture as (we set $\hbar=1$)
\begin{equation}\label{H real space}
	\begin{aligned}
		H=&\sum_n
		\frac{1}{\Delta}\left( \Omega_{c_1}^2+\Omega_{f_1}^2+2\Omega_{c_1}\Omega_{f_1}\cos{k_0x_n} \right)  |e_n\rangle\langle e_n| \\
		&+\sum_n\frac{1}{\Delta}\left( \Omega_{c_2}^2+\Omega_{f_2}^2+2\Omega_{c_2}\Omega_{f_2}\cos{k_0x_n} \right) |m_n\rangle\langle m_n| \\
		&-\sum_n\Delta_{ps} ( |e_n\rangle\langle e_n| + |m_n\rangle\langle m_n| ) - \Delta_p |f_n\rangle\langle f_n| \\
		&+\sum_n [ \frac{1}{\Delta}( \Omega_{c_1}\Omega_{c_2}
		 +\Omega_{f_1}\Omega_{f_2} +\Omega_{c_2}\Omega_{f_1}e^{i k_0x_n} \\
		 &+\Omega_{c_1}\Omega_{f_2}e^{-ik_0x_n}) |e_n\rangle\langle m_n|
		+\Omega_p e^{-i\mathbf{k_p\cdot r_n}}|g_n\rangle\langle f_n| \\
		&+\Omega_s e^{-i\mathbf{k_s\cdot r_n}}|e_n\rangle\langle f_n| + h.c. ]
	\end{aligned}
\end{equation}
where $k_0=|\mathbf{k_c}-\mathbf{k_f}|$ and $\Delta_{ps}=\Delta_p-\Delta_s$. $\mathbf{r_n}$ labels the coordinate of the n-th atom and $x_n$ is the $x$-component of
$\mathbf{r_n}$, where the direction of the $x$-axis is determined by $\mathbf{k_c-k_f}$.
$|e(g,f,m)_n\rangle$ represents the state $|e(g,f,m)\rangle$ of the atom located at $\mathbf{r_n}$.

By introducing the following timed Dicke states
$|f_j\rangle=1/\sqrt{N}\sum_n e^{i[\mathbf{k_p}+j(\mathbf{k_c-k_f})] \cdot\mathbf{r_n}}|f_n\rangle$,  $|e_j\rangle=1/\sqrt{N}\sum_n e^{i[\mathbf{k_p}-\mathbf{k_s}+j(\mathbf{k_c-k_f})]\cdot\mathbf{r_n}}|e_n\rangle$,
$|m_j\rangle=1/\sqrt{N}\sum_n e^{i[\mathbf{k_p}-\mathbf{k_s}+(j-1)(\mathbf{k_c-k_f})]\cdot\mathbf{r_n}}|m_n\rangle$ with $N$ being the total number of atoms, the Hamiltonian in Eq.~\eqref{H real space} can be transformed to a tight-binding model, $H=H_s+H_p$, where
\begin{equation}\label{H momentum space}
	\begin{aligned}
		H_s=&\sum_j \varepsilon_e |e_j\rangle\langle e_j| + \varepsilon_m|m_j\rangle\langle m_j|
		+ \sum_j [t_0|m_j\rangle\langle e_j| \\
		&+t_1|m_{j+1}\rangle\langle e_j|+t_2 |m_{j+2}\rangle\langle e_j|
		+t_3|e_{j+1}\rangle\langle e_j| \\
		&+t_4|m_{j+1}\rangle\langle m_j|+h.c.]
	\end{aligned}
\end{equation}
and $H_p=(\sqrt{N}\Omega_p |g\rangle\langle f_0|+\sum_j \Omega_s |e_j\rangle\langle f_j|+h.c.)
 - \sum_j \Delta_p|f_j\rangle\langle f_j|+\Delta_{ps}(|e_j\rangle\langle e_j|+|m_j\rangle\langle m_j|)$, where  $\varepsilon_e=(\Omega_{c_1}^2+\Omega_{f_1}^2)/\Delta$, $\varepsilon_m=(\Omega_{c_2}^2+\Omega_{f_2}^2)/\Delta$, $t_0=\Omega_{f_1}\Omega_{c_2}/\Delta$, $t_1=(\Omega_{c_1}\Omega_{c_2}+\Omega_{f_1}\Omega_{f_2})/\Delta$, $t_2=\Omega_{c_1}\Omega_{f_2}/\Delta$, $t_3=\Omega_{c_1}\Omega_{f_1}/\Delta$, $t_4=\Omega_{c_2}\Omega_{f_2}/\Delta$ and $|g\rangle=|g_1,g_2,...,g_N\rangle$.

As shown in Fig.~\ref{fig setup}(b), $t_0$ and $t_1$ describe the hoppings between nearest-neighbor sites within and inbetween the unit cells, respectively. $t_2$, $t_3$ and $t_4$ stands for the distinct long-range hopping amplitudes, respectively. More interestingly, in our proposed five-level M-type scheme, all the hopping amplitudes are highly tunable, which can be achieved through varying the Rabi frequencies and detunings. Therefore, various types of Su-Schrieffer-Heeger (SSH) model, including the standard SSH and extended SSH model \cite{su1979solitons,rice1982elementary,creutz1999end,li2015winding,velasco2017realizing}, can be realized. To gain more insight, we rewrite $H_s$ in the real space \cite{wang2020synthesized,cai2019experimental} as

\begin{figure}[t]
	\centering
	\includegraphics[width=0.54\textwidth]{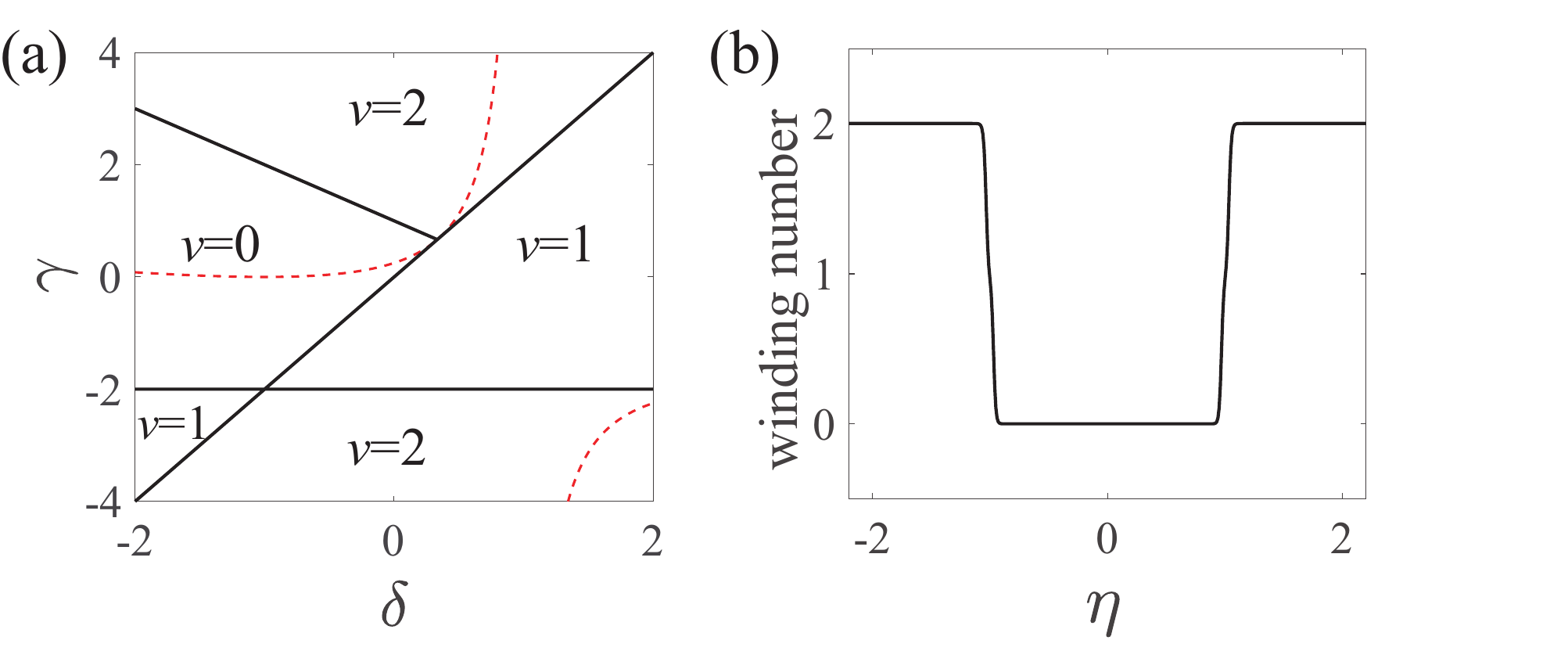}
	\caption{
		(a) The topological phase diagram of the model Hamiltonian in Eq.\eqref{final Hs}.  Since here $\Omega/{\bar \Omega}\equiv\eta$ is the allowed tuning parameter, the model in Eq.\eqref{final Hs} can approach the regime along the red line.
		(b) The topological phase transition between topological trivial regime with winding number $\nu=0$ and topological non-trivial regime with winding number $\nu = 2$ when varying $\eta$.
	}
	\label{fig SL}
\end{figure}

\begin{equation}\label{H dual space}
	\tilde{H_s}= h_x\sigma_x+h_y\sigma_y+h_z\sigma_z+h_0I
\end{equation}
where $h_x=t_0+t_1\cos \tilde{k}_x+t_2\cos 2\tilde{k}_x$, $h_y=t_1\sin \tilde{k}_x+t_2\sin 2\tilde{k}_x$, $h_z=(\varepsilon_e-\varepsilon_m)/2+(t_3-t_4)\cos \tilde{k}_x$,  $h_0=(\varepsilon_e+\varepsilon_m)/2+(t_3+t_4)\cos \tilde{k}_x$ and $\tilde{k}_x=k_0x$. $I$ is the unit matrix and $\mathbf{\sigma}$ is the Pauli's matrix.
Through diagonalizing Eq.~\eqref{H dual space}, the band
dispersion $E^{s}(\tilde{k}_x)$ can be obtained as $E^s_+=\frac{1}{\Delta}\sum_{i=1,2} \Omega_{c_i}^2+\Omega_{f_i}^2+2\Omega_{c_i}\Omega_{f_i}\cos \tilde{k}_x $ and $E^s_-=0$.
Surprisingly, it is shown that the corresponding energy bands exhibit that there is a robust flat band for any Rabi frequency and detuning in our five-level M-type scheme.
To further consider realizing a two-band 1D Z-type topological system from the model in Eq.~\eqref{H dual space}, we consider tuning the Rabi frequencies satisfying the following relations $\Omega_{c_1}=\Omega_{f_2}\equiv \Omega$ and $\Omega_{c_2}=\Omega_{f_1}\equiv \bar{\Omega}$.
Therefore $\tilde{H_s}$ in Eq.~\eqref{H dual space} can be simplified as
\begin{equation}\label{final Hs}
	\tilde{H_s}^{\prime}=h_x\sigma_x+h_y\sigma_y+h_0I
\end{equation}
where $h_x=(\bar{\Omega}^2+2\Omega\bar{\Omega}\cos \tilde{k}_x+\Omega^2\cos 2\tilde{k}_x)/\Delta$, $h_y=(2\Omega\bar{\Omega}\sin \tilde{k}_x+\Omega^2\sin 2\tilde{k}_x)/\Delta$, $h_0=(\Omega^2+\bar{\Omega}^2+2\Omega\bar{\Omega}\cos \tilde{k}_x)/\Delta$. The corresponding band structure can be determined by $E^s_{+}= 2(\Omega^2+\bar{\Omega}^2+2\Omega\bar{\Omega}\cos \tilde{k}_x)/\Delta$ and $E^s_{-}=0$.
From Eq.~\eqref{final Hs}, it is shown that $\Omega/{\bar{\Omega}}\equiv\eta$ is the allowed tuning parameter here. Therefore, the Hamiltonian parameters $\delta \equiv (t_1-t_0)/(t_0+t_1)$ and $\gamma \equiv 2t_2/(t_0+t_1)$ are constrained by the following relation $\delta/\gamma=(2\eta-1)/2\eta^2$.
As shown in Fig.\ref{fig SL}(a), such a constrain will effect the topological properties of the system.

 \begin{figure}[t]
 	\centering
 	\includegraphics[width=0.5\textwidth]{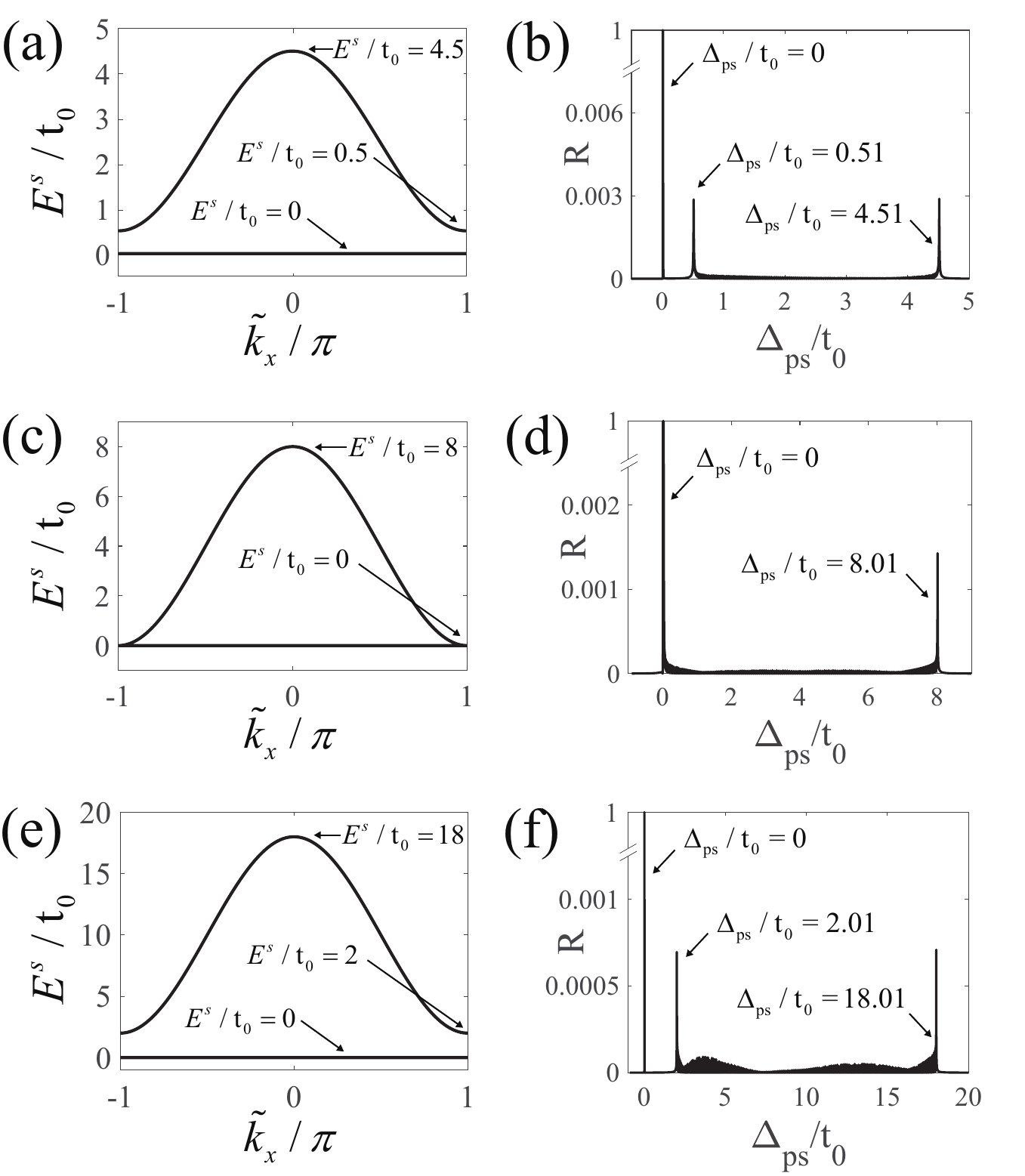}
 	\caption{
 		Energy spectra of the Hamiltonian $\tilde{H}'_s$. (left column) vs. superradiance emission spectra (right column) for different $\eta$: (a) $\eta=0.5$, (b) $\eta=1.0$ and (c) $\eta=2$. It is shown that both the top and bottom of each energy band of $\tilde{H}'_s$. can be determined by the locations of the peaks of the intensity of superradiance emission.
 	}
 	\label{fig band}
 \end{figure}

Since the system belongs to 1D Z-type, the typological nature can be characterized by  the winding number defined as $\nu=\frac{1}{2\pi}\oint_C\frac{h_xdh_y-h_ydh_x}{h_x^2+h_y^2}$, where $C$ is a close loop with $\tilde{k}_x$ varying from 0 to $2\pi$. We find that the model in Eq.~\eqref{final Hs} can approach the topological regime along the red line in Fig.~\ref{fig SL}(a). There is a topological phase transition from the topological trivial regime with $\nu = 0$ to the topological non-trivial regime with $\nu = 2$ when varying $\eta$ as shown in Fig.~\ref{fig SL}(b). More interestingly, we find that when the system in the topological regime, the flat band also shows the non-trivial topological property, which is characterized by a higher winding number $w=\frac{1}{\pi}\int_0^{2\pi} d\tilde{k}_x \langle \psi |i\partial_{\tilde{k}_x} |\psi\rangle$ being 2, where $|\psi\rangle$ is eigen-state of Eq.\eqref{final Hs} corresponding to the flat band.

\section{superradiance spectra}
The above topological phase transition can be detected through the superradiance spectra, which will reflect the changes in the band structure \cite{wang2015superradiance,chen2018experimental,wang2020synthesized,he2021flat,mao2022measuring}.
Distinct from the spontaneous radiation, superradiance emission is a collective effect and generates a stronger directional radiance.
By setting the phase matching condition as  $|\mathbf{k_p}+\mathbf{k_c}-\mathbf{k_f}|=|\mathbf{k_p}|$, superradiance photon will be emitted along the direction determined by $\mathbf{k_p}+\mathbf{k_c}-\mathbf{k_f}$, which is indicated by the dashed line in Fig.~\ref{fig setup}(a).
The intensity of superradiance emission can be calculated by solving the following  master equation
\begin{equation}\label{master eq}
	-i\left[H,\rho\right]+\sum_\alpha \Gamma_\alpha \left( L_\alpha\rho L_\alpha^\dagger -\frac{1}{2}\{ L_\alpha^\dagger L_\alpha, \rho \} \right) =0
\end{equation}
where $\alpha$ runs over $p,s,c,f$ and $\Gamma_\alpha$ describes the corresponding transition decay rate. $L_\alpha$ is the Linblad operator defined as $L_p=|g\rangle\langle f|$, $L_s=|e\rangle\langle f|$, $L_c=|e\rangle\langle d|$ and $L_f=|m\rangle\langle d|$, respectively. Since both the probe and signal fields are weak, atoms are mainly populated at the ground state $|g\rangle$. Then, the density matrix can be expressed as $\rho=|g\rangle\langle g|+\sum_{j}(A_j|e_j\rangle\langle g|+\sum_jB_j|m_j\rangle\langle g|+\sum_{j}C_j|f_j\rangle\langle g|+h.c.)$. Substituting the above expression of density matrix into Eq.\eqref{master eq}, we obtain the following equation for the steady state
\begin{widetext}
	\begin{equation}\label{big matrix}
		\begin{pmatrix}
			& ... &  &  &  &  &  &   \\
			t_0 & \Delta_2-\Delta_{ps} & 0 & t_4 & 0 & 0 &0 &  \\
			t_3 & 0 & \Delta_1-\Delta_{ps} & t_0 & t_3 & t_1 & 0 & t_2  \\
			t_1 & t_4 & t_0 & \Delta_2-\Delta_{ps} & 0 & t_4 & 0 & 0  \\
			0 & 0 & t_3 & 0 & \Delta_1-\Delta_{ps} & t_0 & t_3 & t_1  \\
			t_2 & 0 & t_1 & t_4 & t_0 & \Delta_2-\Delta_{ps} & 0 & t_4  \\
			& 0 & 0 & 0 & t_3 & 0 & \Delta_1-\Delta_{ps} & t_0   \\
			&  &  &  &  &  & ... &
		\end{pmatrix}
		\begin{pmatrix}
			...\\
			B_{-1} \\
			A_{0} \\
			B_{0}\\
			A_1\\
			B_1\\
			A_2\\
			...
		\end{pmatrix}
		=
		\begin{pmatrix}
			...\\
			0\\
			-\frac{\Omega_p\Omega_s}{\Delta_p+i\Gamma} \\
			0\\
			0\\
			0\\
			0\\
			...
		\end{pmatrix}
	\end{equation}
\end{widetext}
 where $\Delta_1=\Delta_2+|\Omega_s|^2/(\Delta_p+i\Gamma)$, $\Delta_2=(\Omega^2+\bar{\Omega}^2)/\Delta$, and $\Gamma$ is the decay rate of $|f\rangle$. Note that under the condition $\Omega_p \ll \Omega,\bar{\Omega}$, it is found that $C_j$ depends linearly on $A_j$ through the relation $C_j=(\Omega_sA_j+\Omega_p\delta_{j,0})/(\Delta_p+i\Gamma)$.
 Through solving the density matrix for the steady state from Eq.\eqref{big matrix},  the electric polarization intensity of atoms can be calculated via the following relation  $\mathbf{P}=-e\mathrm{Tr}[U^\dagger\rho U\mathbf{r}]=\sum_\alpha\mathbf{P_\alpha}\exp(-i\omega_\alpha t)$, where $U=\exp(-iH_0t)$ with $H_0=-\sum_j \omega_p|f_j\rangle\langle f_j|+(\omega_p-\omega_s)(|e_j\rangle\langle e_j|+|m_j\rangle\langle m_j|)$. $\omega_\alpha$ is the frequency of corresponding field and $P_\alpha$ is the polarization intensity excited by corresponding field, respectively. Then, the susceptibility of atomic medium can be obtained via
 \begin{equation}\label{key}
 	\chi=P_p/\varepsilon_0E_p e^{i\mathbf{k_p\cdot r}}=\sum_j \chi_j e^{ij(\mathbf{k_c-k_f})\cdot\mathbf{r}}
 \end{equation}
where $\chi_j=\mu_{eg}C_j/\varepsilon_0 E_pV$ and $\mu_{eg}$ is the electric dipole matrix element related to the atomic transition between $|e\rangle$ and $|g\rangle$. $V$ is the volume of atom ensemble and $\varepsilon_0$ is permittivity of vacuum. The reflectivity $R$ can be calculated through the following equations \cite{chen2018experimental,cai2019experimental}
\begin{equation}\label{key}
	\begin{aligned}
		\partial_xE_p & =-\beta_0E_p + i\kappa_{-1}e^{-ik_0x} E_r\\
		\partial_xE_r & =\beta_0E_r - i\kappa_{+1}e^{ik_0x} E_p
	\end{aligned}
\end{equation}
where $\beta_0=\frac{\nu_p^2}{2k_pc^2\sin\frac{\theta}{2}}\mathrm{Im}(\chi_0)$ and $\kappa_{\pm 1}=\frac{\nu_p^2}{2k_pc^2\sin\frac{\theta}{2}}\chi_{\pm 1}$. $E_p$ is the field amplitude of the incident probe beam and $E_r$ is the field amplitude of the scattered beam, i.e. superradiance emission. The reflectivity $R=|E_r(0)/E_p(0)|^2$, under the boundary condition $E_p(0)=E_0$, $E_r(L)=0$, can be solved though the following relation
\begin{equation}\label{R}
	R=\left|\frac{\kappa_{+1}(e^{-\lambda L}-e^{\lambda L})}{(\beta-\lambda)e^{-\lambda L}-(\beta+\lambda)e^{\lambda L}} \right|^2
\end{equation}
where $\beta=\beta_0-\frac{i}{2}k_0$ and $\lambda=\sqrt{\beta^2+\kappa_{+1}\kappa_{-1}}$. $L$ is the length of the system  along the $x$-axis.
From Eq.\eqref{R}, it is found that $R \propto A_1$. And $A_1$ can be obtained from Eq.\eqref{big matrix}, which satisfies the relation $A_1(\Delta_{ps}) \propto \mathrm{Dos}(E^s)|_{E^s=\Delta_{ps}}$ with $\mathrm{Dos}(E^s)$ being the density of state
of the band spectra $E^s(\tilde{k}_x)$ of the Hamiltonian $\tilde{H}'_s$. Therefore,  when $\Delta_{ps} $ approaching the energy where $\mathrm{Dos}(E^s)$ diverges, there should be a peak at the intensity of superradiant emission.
As shown in Fig.~\ref{fig band}, all the peaks of the intensity of superradiant emission are located at the saddle point of the energy spectra $E^s(\tilde{k}_x)$. The top and bottom of each energy band of $H_s$ can thus be determined.  As shown in Fig.~\ref{fig band} from top line to bottom line, when varing $\eta$, there is a topological phase transition. At the transition point, the system become gapless and there is only one energy band. It can be detected from counting the peaks of the intensity of superradiant emission. As shown in Fig.~\ref{fig band}(d), there are only two peaks in $R$, indicating that the energy gap is closed and the topological phase transition occurs. Therefore, the topological phase diagram as shown in Fig.~\ref{fig SL}(b) can be determined through measuring the superradiance spectra.

\section{Conclusion}
In summary, we propose a five-level M-type scheme in atomic ensembles to induce a 1D bipartite superradiance lattice in momentum space.
Such a lattice shows great tunability of changing both the nearest-neighbor and long-range hopping amplitude through varying the Rabi frequencies and detunings.  Various types of SSH models can thus be achieved. A flat band with higher winding number can be achieved. We also proposed that the superradiance spectra can be utilized as a tool for experimental detection.  Our proposal would provide a promising approach to explore the new physics in the flat bands with higher topological number.

\section{Acknowledgement}
This work is supported by the National Key R$\&$D Program of China (2021YFA1401700), NSFC (Grants No. 12074305, 12147137, 11774282), the National Key Research and Development Program of China (2018YFA0307600), Xiaomi Young Scholar Program (S. L., R. T.,H. W., M. L. and B. L.), and NSFC (Grant No. 12004229) (L. C.). We also thank the HPC platform of Xi'An Jiaotong University, where our numerical calculations was performed.

\bibliographystyle{apsrev}
\bibliography{SL-ref}
	
\end{document}